\documentclass[prb,preprint]{revtex4-1}
\usepackage{graphicx}
\usepackage{amssymb}
\usepackage{amsmath}
\usepackage{chngcntr}
\usepackage{listings}
\usepackage{xcolor}

\usepackage[T2A,T1]{fontenc}
\DeclareSymbolFont{cyrillic}{T2A}{cmr}{m}{n}
\DeclareMathSymbol{\Sha}{\mathalpha}{cyrillic}{216}

\begin{document}
\title{Which part of a chain breaks}
\author{Seung Ki Baek}
\email{Electronic mail: seungki@pknu.ac.kr}
\affiliation{Department of Physics, Pukyong National University, Busan 48513,
Korea}

\date{\today}

\begin{abstract}
This work investigates the dynamics of a one-dimensional homogeneous harmonic
chain on a horizontal table.
One end is anchored to a wall; the other (free) end is pulled by
external force. A Green's function is derived to calculate the response to a
generic pulling force. As an example, I assume that the magnitude of the
pulling force increases with time at a uniform rate $\beta$. If the number of
beads and springs used to model the chain
is large, the extension of each spring takes a
simple closed form, which is a piecewise-linear function of time.
Under an additional assumption that a spring breaks when its extension exceeds a
certain threshold, results show that for large $\beta$ the spring breaks near
the pulling
end, whereas the breaking point can be located close to the wall by choosing
small $\beta$. More precisely, the breaking point moves back and forth along
the chain as $\beta$ decreases, which has been called ``anomalous'' breaking in
the context of the pull-or-jerk experiment. Although the experiment has been
explained in terms of inertia, its meaning can be fully
captured by discussing the competition between intrinsic and extrinsic time
scales of forced oscillation.
\end{abstract}

\pacs{01.30.mp,62.20.M-,63.20.-e}

\maketitle

\section{Introduction}

\begin{figure}
\includegraphics[width=0.15\textwidth]{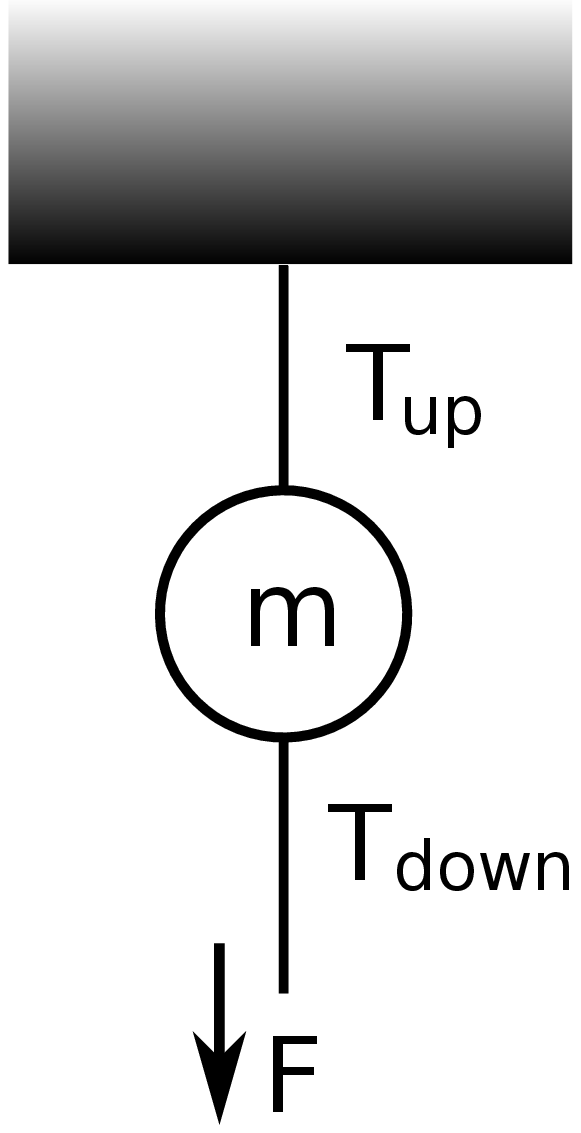}
\caption{Typical configuration of the pull-or-jerk experiment consisting of
two elastic strings and a ball of mass $m$. The upper string has tension
$T_{\text{up}}$. The lower one has $T_{\text{down}}$ and a time-dependent force
$F$ is exerted at its end.
}
\label{fig:ball}
\end{figure}

The pull-or-jerk (or ``inertia ball'') experiment is commonly used to
demonstrate
Newton's laws of motion: As depicted in Fig.~\ref{fig:ball}, a ball of mass
$m$ is hung from a ceiling by a string, whose tension is denoted as
$T_\text{up}$. Another string is attached at the bottom of the ball, and its
tension is denoted as $T_\text{down}$. A downward ``jerky'' force $F$
is exerted at the end of the lower string, and the question is which string
breaks. We know from our experience that the answer depends on how quickly the
magnitude of the force changes: If the force
suddenly increases, the lower string
breaks. If, on the other hand, the force increases slowly, then the upper string
breaks. This
phenomenon is sometimes explained as due to the inertia of the ball, described
as a ``tendency to resist changes in motion''~(see, e.g.,
Refs.~\onlinecite{hewitt2009conceptual} and \onlinecite{iop}).
However, this does not clearly
answer why the outcome varies with the jerkiness of the driving force, if all we
know about mass is that it is a constant of motion independent of any specific
dynamic process. Moreover, this experiment has a counter-intuitive aspect,
which may be completely baffling unless one makes good sense of inertia:
Suppose that the force has constant jerkiness $\alpha$ so that $F =
\alpha t$ for $t > 0$. By approximating each string as a spring with force
constant $K$, one can solve the equation of motion to
obtain\cite{heald1996string,caplan2004ye}
\begin{equation}
\left\{
\begin{array}{lcl}
T_\text{up} &=& \alpha t - \alpha\omega_0^{-1}\sin \omega_0 t\\
T_\text{down} &=& \alpha t,
\end{array} \right.
\label{eq:heald}
\end{equation}
where $\omega_0 \equiv \sqrt{K/m}$ and the gravitational force is neglected.
Interested readers are referred to Ref.~\onlinecite{shima2014analytic}
for a thorough analysis of this system.
A simple assumption on the failure behavior is that the string
breaks when $T$ exceeds a certain threshold, say, $T_c$. According to
Eq.~\eqref{eq:heald}, the lower string will break if $\alpha > \alpha_c \equiv
\omega_0 \pi^{-1} T_c$, which is consistent with our understanding. At the same
time, it also predicts the existence of ``anomalous'' breaking, which means that
the force can break the lower string with even smaller jerkiness $\alpha \in
\left[ \frac{1}{3} \alpha_c, \frac{1}{2} \alpha_c \right]$. In this sense, which
string breaks is not really a matter of pull or jerk.

\begin{figure}
\includegraphics[width=0.9\textwidth]{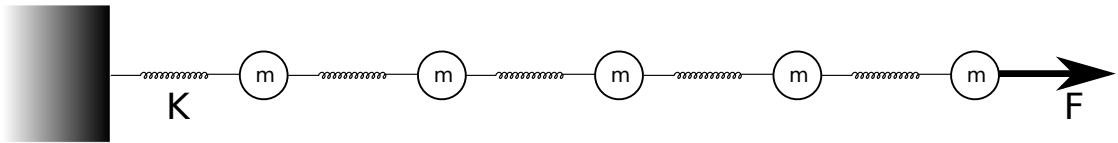}
\caption{Schematic representation of a harmonic chain with $N=5$ beads.
Every bead has mass $m$, whereas the spring is massless. The spring constant
is $K$ for every spring.
The leftmost spring is fastened to the wall whose displacement is fixed as
$y_0 = 0$, and the rightmost bead is driven by an external force $F$.
The whole system is placed on a horizontal table
so that the gravitational force does not enter the equation of motion.
}
\label{fig:harmonic}
\end{figure}

A natural question would be how this analysis generalizes to a chain of many
beads and springs,\cite{shima2014analytic} which I wish to address in this
work.
A harmonic chain is a useful starting point to investigate properties of a
macroscopic system near equilibrium.
It has been used to understand basic statistical properties of
solids such as heat capacity\cite{swendsen2012introduction} and thermal
conductivity\cite{rieder1967properties,florencio1985exact,lee1989dynamic,hu2000heat,dhar2001heat,bonetto2004fourier}
and the breaking strength of a polymer
chain.\cite{doerr1994breaking,lee2009thermal} A ladder
of resistively and capacitively shunted Josephson junctions can also be
approximated by a harmonic chain through the mapping to a locally coupled
one-dimensional Kuramoto
model\cite{wiesenfeld1996synchronization,wiesenfeld1998frequency,daniels2003synchronization}
when phase differences are small. Specifically,
I will consider a harmonic chain consisting of identical beads and
springs on a horizontal table as depicted in Fig.~\ref{fig:harmonic}. One end of
the chain is fixed to the wall, and the other end is driven by a time-varying
external force $F$.
The primary goal of this paper is to give students
a precise picture of this general many-body pull-or-jerk experiment.
Following Ref.~\onlinecite{heald1996string},
the spring is assumed to have a threshold of
deformation above which it ceases to obey Hooke's law. The question is
which spring is the earliest that reaches the threshold under a given external
force, and this spring will be regarded as a breaking point of the chain.
My finding is that the oscillatory motion in Eq.~\eqref{eq:heald} manifests
itself as a wave traveling across this many-body system, implying that one
should consider two competing time scales, one for external driving and the
other for internal wave dynamics, to understand the failure behavior.
This study can also be thought of as an advanced exercise for physics majors
because a harmonic chain is a representative mechanical example that is
analytically soluble by means of undergraduate-level
mathematics.\cite{boas2006mathematical}

This work is organized as follows: In Sec.~\ref{sec:model}, I calculate the
Green's
function for the model system by solving the full equation of motion. It gives
an approximate formula which holds in the continuum limit. The case of a
linearly increasing force is then investigated in Sec.~\ref{sec:application}
under the assumption that friction is negligibly small. The analytic result is
compared with numerical integration of the equations of motion. I discuss
implications of the observed behavior and conclude this work in
Sec.~\ref{sec:conclusion}. A sample Python code is provided in the Appendix.

\section{Model}
\label{sec:model}

Consider longitudinal waves on a harmonic chain consisting of beads
and springs. Each bead has mass $m$ and every spring has the same spring
constant $K$, and the square root of their ratio is defined as $\omega_0 \equiv
\sqrt{K/m}$. The number of beads is $N$, and their equilibrium positions in the
absence of external force is denoted by $x_j = jl$, where $l$ means the
equilibrium length of the spring ($j=1,\ldots,N$). The total length of the
system in equilibrium therefore equals $x_N = Nl \equiv L$.
The displacement of the $j$th bead from its equilibrium position is
denoted by $y_j$. The number of springs is also
$N$, and the extension of the $j$th spring is $z_j \equiv y_j-y_{j-1}$.
The $N$th bead is pulled by a time-dependent external force $F(t)$, where $t$
denotes time. The equations of motion can thus be written as follows:
\begin{equation}
\left\{
\begin{array}{ll}
m \frac{d^2}{dt^2} {y}_1 =
K (- 2y_{1} + y_{2}) - \Gamma \frac{d}{dt} {y}_1 &\\
m \frac{d^2}{dt^2} {y}_j =
K (y_{j-1} - 2y_{j} + y_{j+1}) - \Gamma \frac{d}{dt} {y}_j &
\mbox{for~}1 < j < N\\
m \frac{d^2}{dt^2} {y}_N =
K (y_{N-1} - y_{N}) - \Gamma \frac{d}{dt} {y}_N + F(t) &
\end{array}
\right.
\label{eq:motion0}
\end{equation}
where $\Gamma$ is a friction coefficient. With
the Kronecker delta $\delta_{k,l}$ and two auxiliary variables $y_0 \equiv 0$
and $y_{N+1} \equiv y_N$, all the above cases can be covered by the following
expression:
\begin{equation}
m \frac{d^2}{dt^2} {y}_j =
K (y_{j-1} - 2y_{j} + y_{j+1}) - \Gamma \frac{d}{dt} {y}_j + F(t) \delta_{j,N},
\label{eq:motion}
\end{equation}
where $j = 1, \ldots, N$. In this notation, $y_0 = 0$ and $y_{N+1} = y_N$ can be
regarded as boundary conditions of Eq.~\eqref{eq:motion}. In particular, $y_0=0$
has direct physical meaning because the wall can be regarded as a fictitious
bead with zero displacement (see Fig.~\ref{fig:harmonic}). In addition, the
system is initially at rest with zero displacements, i.e., $y_j = 0$ and
$d{y}_j/dt = 0$ for every $j$ at $t=0$. In a dimensionless form, the dynamics is
now rewritten as
\begin{equation}
\frac{d^2}{d\tau^2} \psi_j = (\psi_{j-1} - 2\psi_{j} + \psi_{j+1}) - 2\gamma
\frac{d}{d\tau} \psi_j + f(\tau) \delta_{j,N},
\label{eq:dimless}
\end{equation}
where $\tau \equiv \omega_0 t$, $\psi_j \equiv y_j / l$, $\gamma \equiv \Gamma /
(2m\omega_0)$, and $f \equiv F / (K l)$. It is convenient to choose $f(\tau) =
u(\tau)$ and solve Eq.~\eqref{eq:dimless} for $\tau > 0$, where $u$ means the
Heaviside step function.
The unknown $\psi_j$ is decomposed into homogeneous and
particular parts, denoted by $\psi_j^{\text{(h)}}$ and
$\psi_j^{\text{(p)}}$, respectively, to have $\psi_j =
\psi_j^{\text{(h)}} + \psi_j^{\text{(p)}}$. It is easy to see that
$\psi_j^{\text{(p)}} = j$ constitutes a particular solution for
$j=1,\ldots,N$ with $\psi_{0}^{\text{(p)}} \equiv 0$ and
$\psi_{N+1}^{\text{(p)}} \equiv \psi_{N}^{\text{(p)}}$. On
the other hand, $\psi_j^{\text{(h)}}$ satisfies the following homogeneous
equation:
\begin{equation}
\frac{d^2}{d\tau^2} \psi_j^{\text{(h)}} = \psi_{j-1}^{\text{(h)}}
- 2\psi_{j}^{\text{(h)}} + \psi_{j+1}^{\text{(h)}}
- 2\gamma \frac{d}{d\tau} \psi_j^{\text{(h)}}
\label{eq:hom}
\end{equation}
with $\psi_0^{\text{(h)}} \equiv 0$ and $\psi_{N+1}^{\text{(h)}} \equiv
\psi_{N}^{\text{(h)}}$. The initial conditions are $\psi_j^{\text{(h)}} =
-j$ and $d\psi_j^{\text{(h)}}/d\tau = 0$ for $j=1,\ldots,N$ at
$\tau=0$.
To construct a solution, one has to choose $\psi_j^{\text{(h)}}
\propto \sin k j$, considering $\psi_0^{\text{(h)}} = 0$.
The other boundary condition $\psi_{N+1}^{\text{(h)}} = \psi_{N}^{\text{(h)}}$
is then rewritten as $\sin k(N+1) = \sin kN$, which
quantizes the wavenumber as
$k_n = (n+\frac{1}{2})\pi/(N+\frac{1}{2})$ with $n=0, 1, \ldots, N-1$.
Note that the resulting
basis functions are orthogonal in the sense that
$\frac{4}{2N+1}\sum_{j=1}^N \sin k_n j \sin k_m j = \delta_{mn}$. With these
basis functions, the homogeneous solution is represented as
$\psi_j^{\text{(h)}} = \sum_{n=0}^{N-1} a_n (\tau)
\sin k_n j$. Substituting this into Eq.~\eqref{eq:hom}, one sees that
the coefficients have to satisfy
\begin{equation}
\frac{d^2}{d\tau^2} a_n + 2\gamma \frac{d}{d\tau} a_n = - \Omega_n^2 a_n
\end{equation}
with $\Omega_n \equiv 2 \sin (k_n/2)$, which is just
the dispersion relation for phonons.
The above differential equation can be
solved by $a_n (\tau) = A e^{\mu_n^+ \tau} + B e^{\mu_n^- \tau}$ with
$\mu_n^\pm \equiv -\gamma \pm \sqrt{\gamma^2 - \Omega_n^2}$ and arbitrary
constants $A$ and $B$. The constants are determined by applying the
orthogonality relation to the initial conditions as follows:
\begin{eqnarray}
a_n(0) &=& A+B = \frac{4}{2N+1} \sum_{j=1}^N (-j) \sin k_n j =
-\frac{c_n}{2\Omega_n^2}\\
\frac{d a_n}{d\tau}(0) &=& A\mu_n^+ + B\mu_n^- = \frac{4}{2N+1} \sum_{j=1}^N 0
\times \sin k_n j =  0,
\end{eqnarray}
where
$c_n \equiv \sin [(1+N)k_n] / \left( N + \frac{1}{2} \right)$.
After some algebra to compute $A$ and $B$ from the above set of equations,
the solution is obtained as the following shifted discrete Fourier series:
\begin{equation}
\psi_j^{\text{(h)}}(\tau) = \sum_{n=0}^{N-1}
\frac{c_n}{\Omega_n^2\sqrt{\gamma^2 - \Omega_n^2}} \left(
\mu_n^- e^{\mu_n^+ \tau} - \mu_n^+ e^{\mu_n^- \tau} \right) \sin k_n j.
\label{eq:exactpsi}
\end{equation}
Using the connection between the Heaviside step function and the Dirac delta
function, i.e., $du/d\tau = \delta(\tau)$, one readily obtains the Green's
function as follows:
\begin{equation}
G_j(\tau) = \frac{d}{d \tau} \psi_j(\tau)
= \sum_{n=0}^{N-1}
\frac{c_n}{\sqrt{\gamma^2 - \Omega_n^2}}
\left( e^{\mu_n^+ \tau} - e^{\mu_n^- \tau} \right) \sin k_n j,
\label{eq:exactgreen}
\end{equation}
which is the response to $f(\tau) = \delta(\tau)$.
Given any $f(\tau)$, the response can
thus be calculated from the following convolution formula:
\begin{equation}
\psi_j(\tau) = \int_0^\tau G_j(\tau') f(\tau-\tau') d\tau'.
\label{eq:exactconv}
\end{equation}

\begin{figure}
\includegraphics[width=0.45\textwidth]{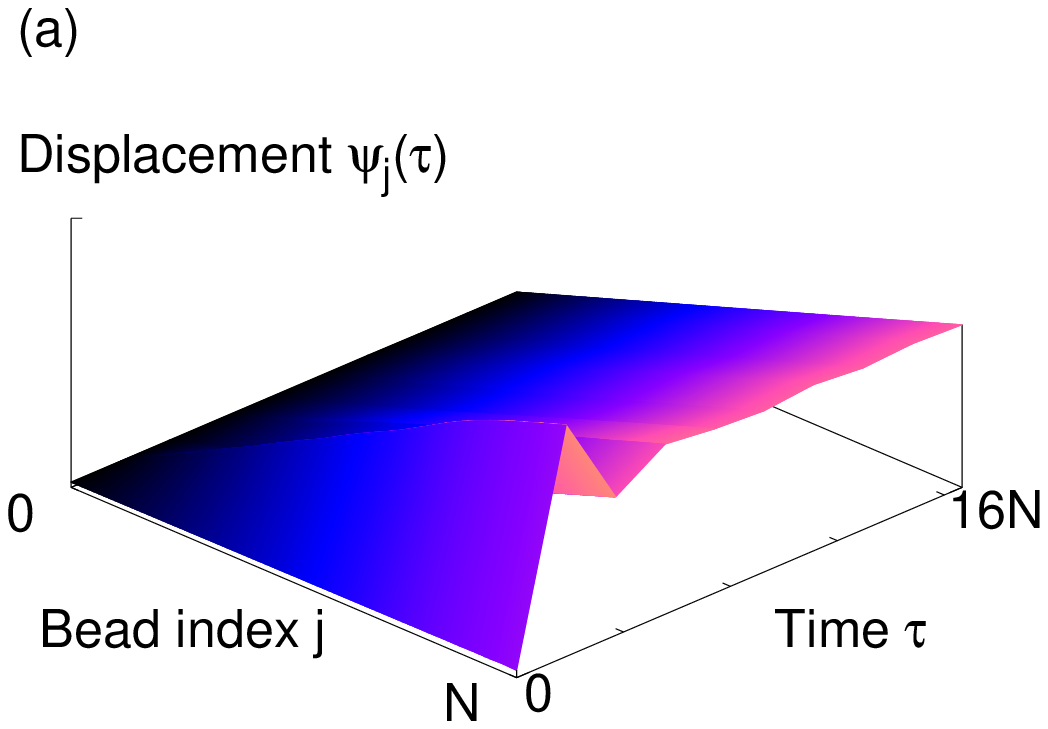}
\includegraphics[width=0.45\textwidth]{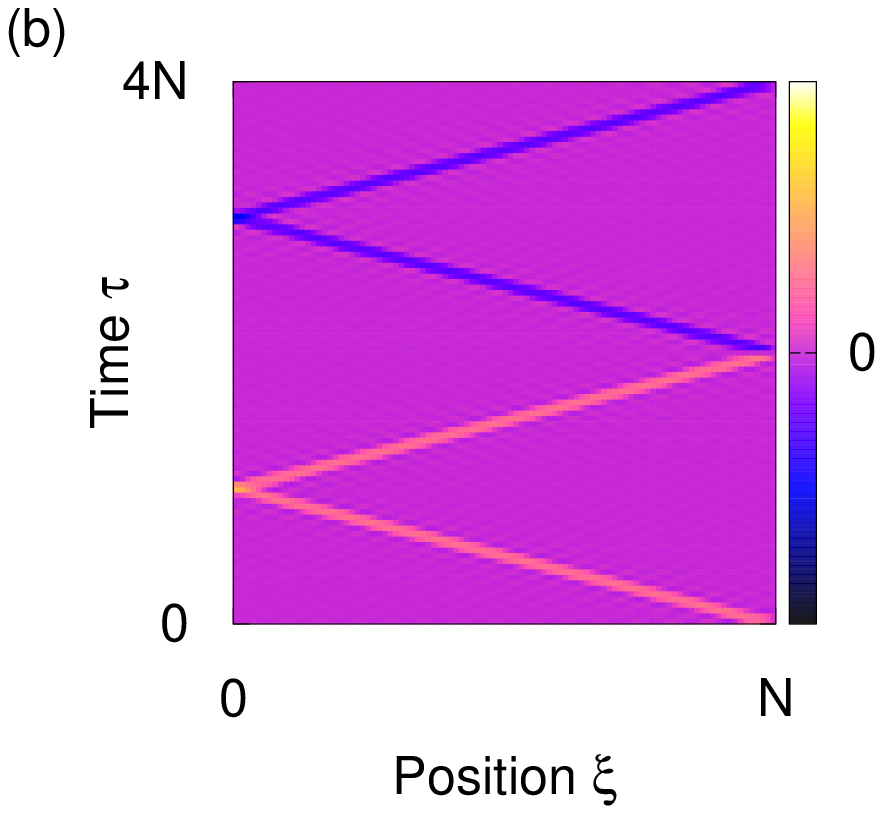}\\
\includegraphics[width=0.45\textwidth]{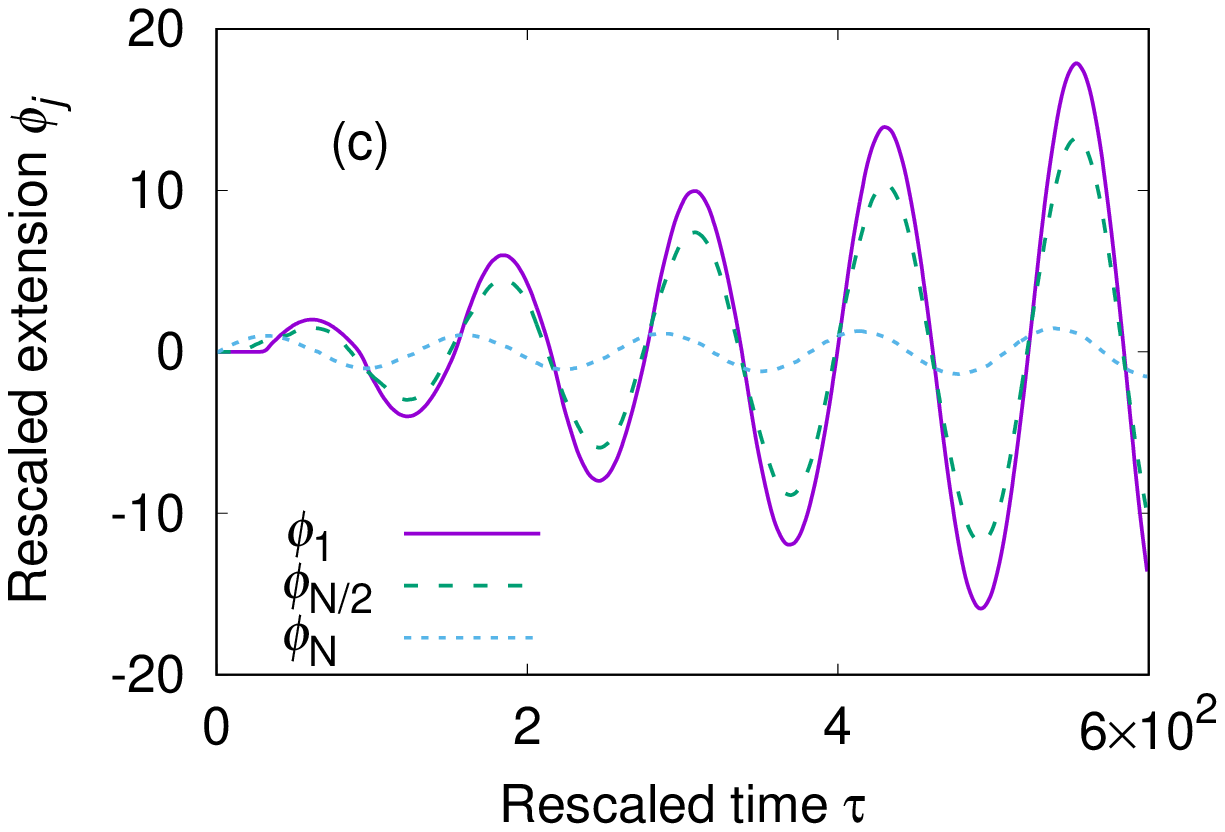}
\includegraphics[width=0.45\textwidth]{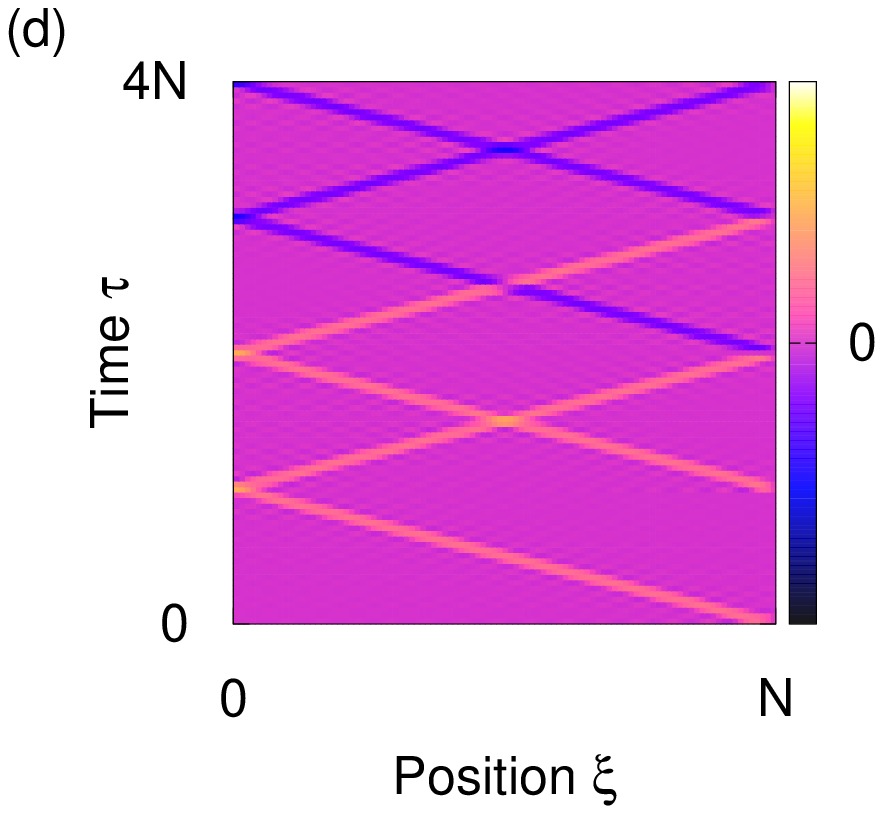}
\caption{(a) Approximate displacement field $\psi(\xi,\tau)$ in
Eq.~\eqref{eq:conv} in response to the Heaviside step function, $f(\tau) =
u(\tau)$, when $N=30$ and $\gamma=10^{-2}$. As time goes by, it converges to the
particular solution, $\psi^{\text{(p)}} (\xi, \tau) = \xi$.
(b) Propagator of Eq.~\eqref{eq:conv2}, $\partial G/\partial \xi$, when
friction is ignored by setting $\gamma=0$.
(c) Extensions of springs under $f(\tau) = \sin(2\pi
\tau/\tau_0)$ with $\tau_0 = 4N$. Equation~\eqref{eq:dimless} is integrated by
the fourth-order Runge-Kutta method (RK4) with $N=30$ and $\gamma = 0$, and the
result agrees well with Eq.~\eqref{eq:sum}. (d) $\phi(\xi,\tau)$ when $f(\tau) =
\delta(\tau) + \delta(\tau-N)$.}
\label{fig:psiu}
\end{figure}

It turns out that the case of $N \gg 1$ greatly simplifies the analysis, making
$k_n \approx \kappa_n \equiv \left( n + \frac{1}{2} \right) \pi / N$,
$\Omega_n \approx \kappa_n$, and $\mu_n^\pm \approx \eta_n^\pm \equiv
-\gamma \pm \sqrt{\gamma^2 - \kappa_n^2}$ for finite $n$.
If a continuous variable $\xi \equiv x/l$ is introduced to replace the integer
index $j$, the Green's function becomes
\begin{equation}
G(\xi,\tau)
\approx \sum_{n=0}^{N-1} \frac{(-1)^n}{N\sqrt{\gamma^2 - \kappa_n^2}}
\left( e^{\eta_n^+ \tau} - e^{\eta_n^- \tau} \right) \sin \kappa_n \xi,
\label{eq:green}
\end{equation}
with the displacement field,
\begin{equation}
\psi(\xi,\tau) = \int_0^\tau G(\xi,\tau') f(\tau-\tau') d\tau',
\label{eq:conv}
\end{equation}
as depicted in Fig.~\ref{fig:psiu}(a).
The rescaled extension of the $j$th spring, $\phi_j \equiv \psi_j -
\psi_{j-1} = z_j/l$, is approximated by $\phi(\xi, \tau) \equiv
\frac{\partial}{\partial \xi} \psi (\xi,\tau)$ evaluated at $\xi=j$. In terms of
Eq.~\eqref{eq:conv}, it is written as
\begin{equation}
\phi(\xi, \tau) = \int_0^\tau \frac{\partial}{\partial \xi}G(\xi, \tau')
f(\tau-\tau') d\tau'.
\label{eq:conv2}
\end{equation}
If $\gamma = 0$ in Eq.~\eqref{eq:green}, the kernel function $\partial G /
\partial \xi$ takes the following form:
\begin{eqnarray}
\frac{\partial G}{\partial \xi} (\xi,\tau)
&\approx& \sum_{n=0}^{N-1} \frac{(-1)^n}{N} 2 \sin \kappa_n \tau \cos \kappa_n \xi\\
&=& \sum_{n=0}^{N-1} \frac{(-1)^n}{N} \left[ \sin \kappa_n (\tau+\xi) + \sin
\kappa_n (\tau - \xi) \right]
\end{eqnarray}
inside the physical region, i.e., $\tau > 0$ and $0 < \xi < N$.
Here, one can verify the following equality:
\begin{equation}
H_N(r) \equiv \sum_{n=0}^{N-1} \frac{(-1)^n}{N} \sin \kappa_n r =
\frac{(-1)^{N-1} \sin \pi r}{2N\cos\frac{\pi r}{2N}}
\end{equation}
by calculating geometric series.
When the argument $r$ is away from $(2p+1)N$ for any integer $p$,
the magnitude of $H_N(r)$ is
small because of $N$ in the denominator. If $r - (2p+1) N = \epsilon \ll 1$,
on the other hand, the cosine in the
denominator behaves linearly as $\epsilon$ varies. It implies
that $H_N$ is well approximated by the normalized sinc function,
${\rm sinc}_\pi (\epsilon) \equiv \sin \pi \epsilon / (\pi \epsilon)$, and
the sign depends on $p$ as follows:
\begin{equation}
H_N \approx (-1)^p {\rm sinc}_\pi (\epsilon).
\end{equation}
To sum up, $H_N(r)$ can be regarded as a train of sinc-typed
impulses at $r=(2p+1)N$. The factor of $(-1)^p$ means that two neighboring
impulses have different signs, so the period of $H_N$ is $4N$ in total.
It would thus be useful to consider a convoluted function $W(r) \equiv {\rm
sinc}_\pi(r) * \Sha_{4N}(r)$, where $\Sha_{4N}(r)$ is the Dirac comb with
periodicity of $4N$. The alternating impulse train is then described as
$H_N(r) \approx W(r-N) - W(r-3N)$ to a good approximation.
Furthermore, one may simply take $W(r) \approx \Sha_{4N}(r)$ because
the convoluted sinc function only modifies the peak shape
without changing the essential physics.
This leads to a particularly
handy formula, $H_N(r) \approx \Sha_{4N}(r-N) - \Sha_{4N}(r-3N)$. Now,
the kernel function simplifies to
\begin{eqnarray}
\frac{\partial G}{\partial \xi} (\xi,\tau) &\approx& H_N(\tau+\xi) +
H_N(\tau-\xi)\\
&\approx& \Sha_{4N}(\tau+\xi-N) - \Sha_{4N}(\tau+\xi-3N)\nonumber\\
&+& \Sha_{4N}(\tau-\xi-N) - \Sha_{4N}(\tau-\xi-3N),
\end{eqnarray}
if $\tau > 0$ and $0 < \xi < N$.
If the Dirac comb is written as an explicit sum of delta
peaks, this can also be expressed as
\begin{eqnarray}
\frac{\partial G}{\partial \xi} (\xi,\tau)
&\approx& \delta(\tau+\xi-N) - \delta(\tau+\xi-3N) + \delta(\tau-\xi-N) -
\delta(\tau-\xi-3N)\nonumber\\
&+& \delta(\tau+\xi-5N) - \delta(\tau+\xi-7N) + \delta(\tau-\xi-5N) -
\delta(\tau-\xi-7N) + \ldots\\
&=& \sum_{\nu=0}^\infty (-1)^\nu \delta[\tau+\xi-(2\nu+1)N]
+ \sum_{\nu=0}^\infty (-1)^\nu \delta[\tau-\xi-(2\nu+1)N].
\label{eq:dg}
\end{eqnarray}
On the $(\xi,\tau)$ plane,
it is basically a pulse propagating back and forth between the two ends of the
chain [Fig.~\ref{fig:psiu}(b)]. The pulse undergoes a phase shift of $\pi$ every
time it hits the free end on the right.
Note that such soliton-like motion is due to the continuum
approximation, in which the wave speed is given independent of the wave number
when $\gamma = 0$. In a finite-sized system, the pulse will
eventually disperse.
Plugging Eq.~\eqref{eq:dg} into Eq.~\eqref{eq:conv2}, one
approximately obtains the extension of the $j$th spring as follows:
\begin{equation}
\phi_j(\tau) \approx \sum_{\nu=0}^{\nu_{\max}^+} (-1)^\nu f[\tau+j-(2\nu+1)N]
+ \sum_{\nu=0}^{\nu_{\max}^-} (-1)^\nu f[\tau-j-(2\nu+1)N],
\label{eq:sum}
\end{equation}
where each of $\nu_{\max}^\pm$ is defined as the greatest integer that makes
positive the argument of every function in the summation.
An example is the periodic driving force $f(\tau) = \sin(2\pi \tau/\tau_0)$ with
$\tau_0 = 4N$. As expected, this induces resonant behavior
[Fig.~\ref{fig:psiu}(c)] because $\partial G/\partial \xi$ has
$4N$-periodicity in time.
It is also clear that one can observe constructive or destructive interference
at a specific spring by sending pulses with an appropriate time interval
[Fig.~\ref{fig:psiu}(d)].

\section{Application}
\label{sec:application}

\begin{figure}
\includegraphics[width=0.45\textwidth]{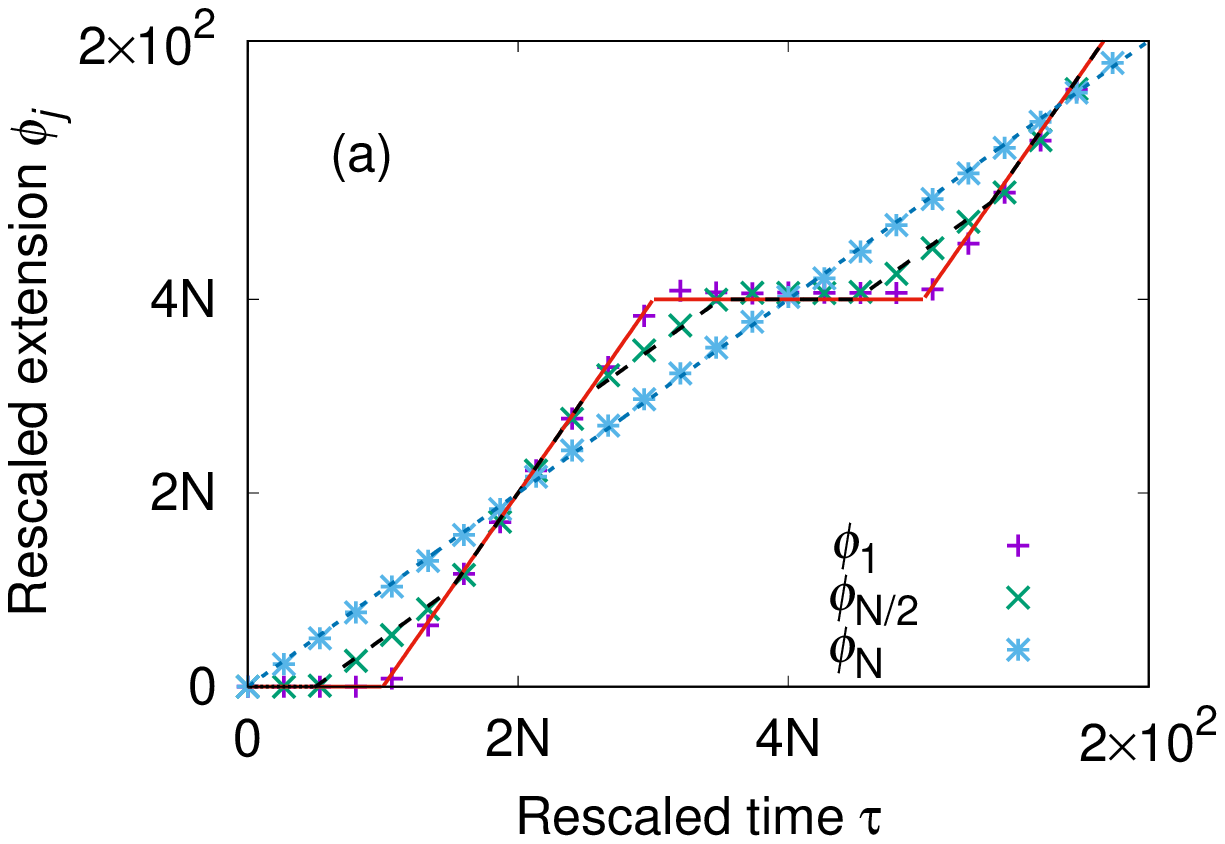}
\includegraphics[width=0.45\textwidth]{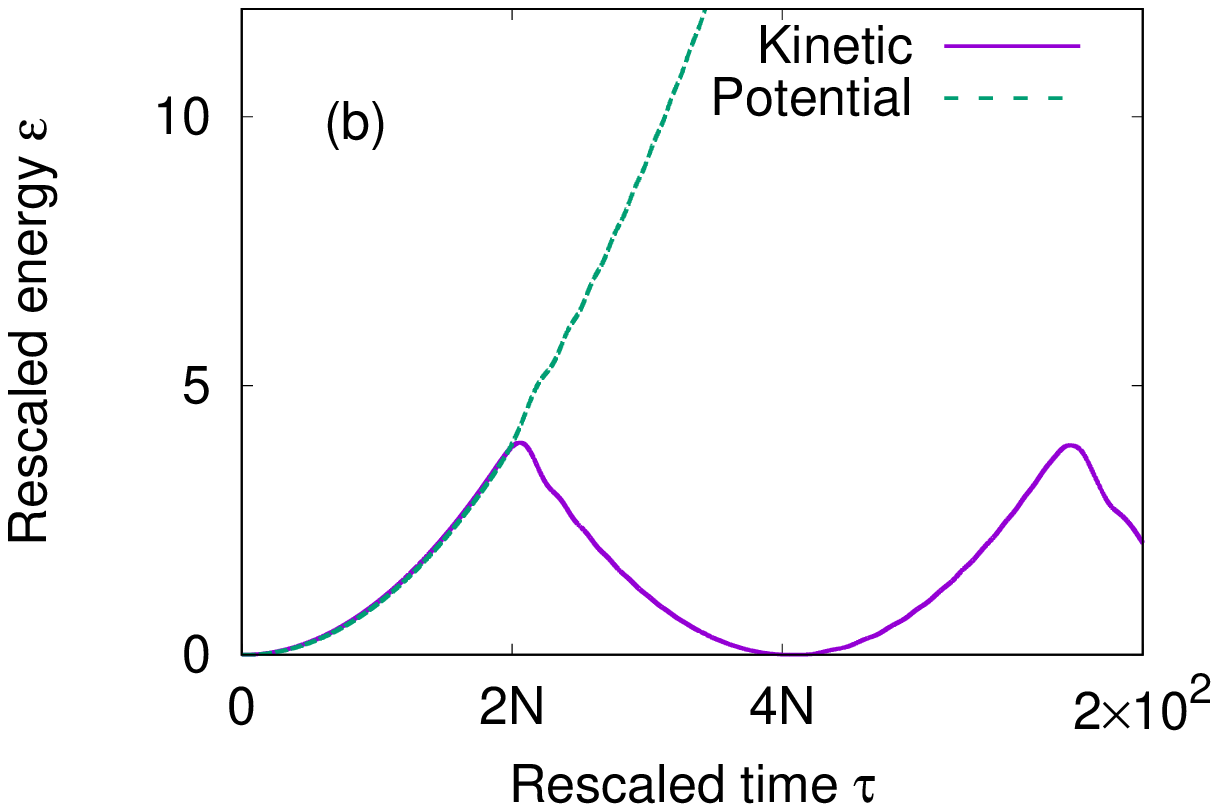}\\
\includegraphics[width=0.45\textwidth]{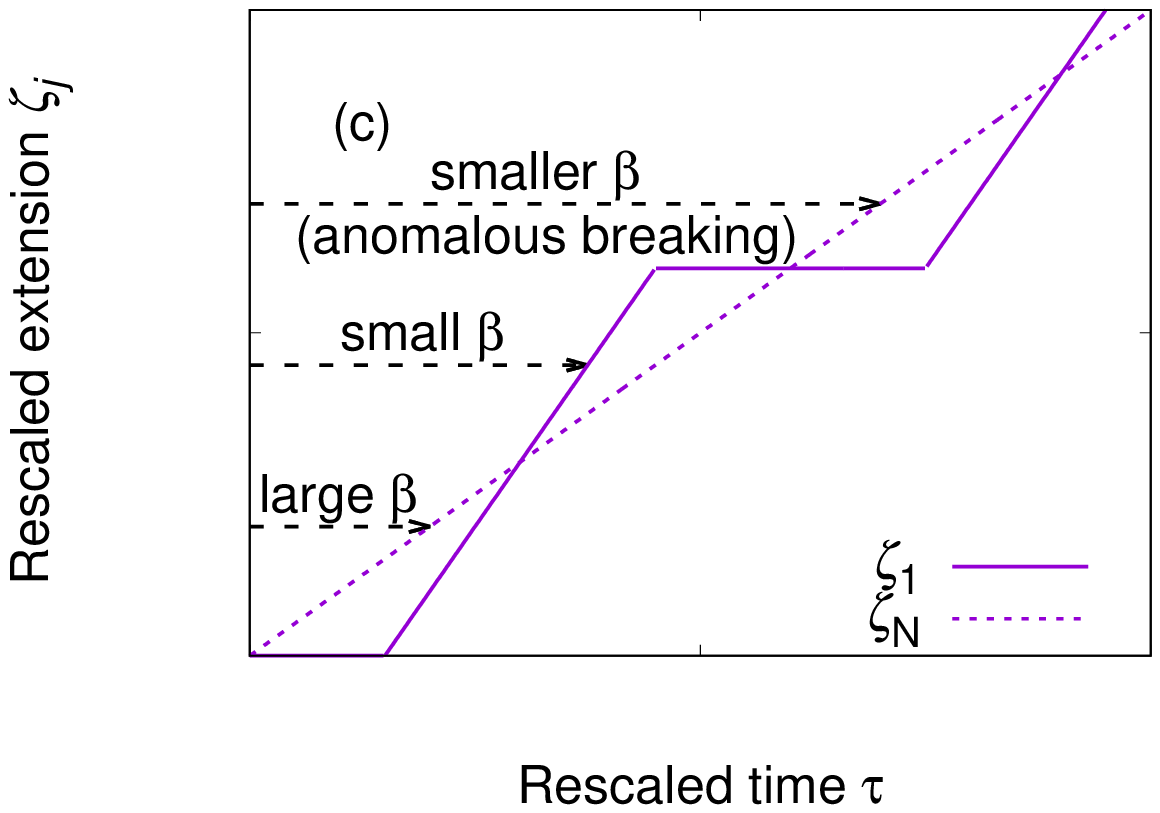}
\includegraphics[width=0.45\textwidth]{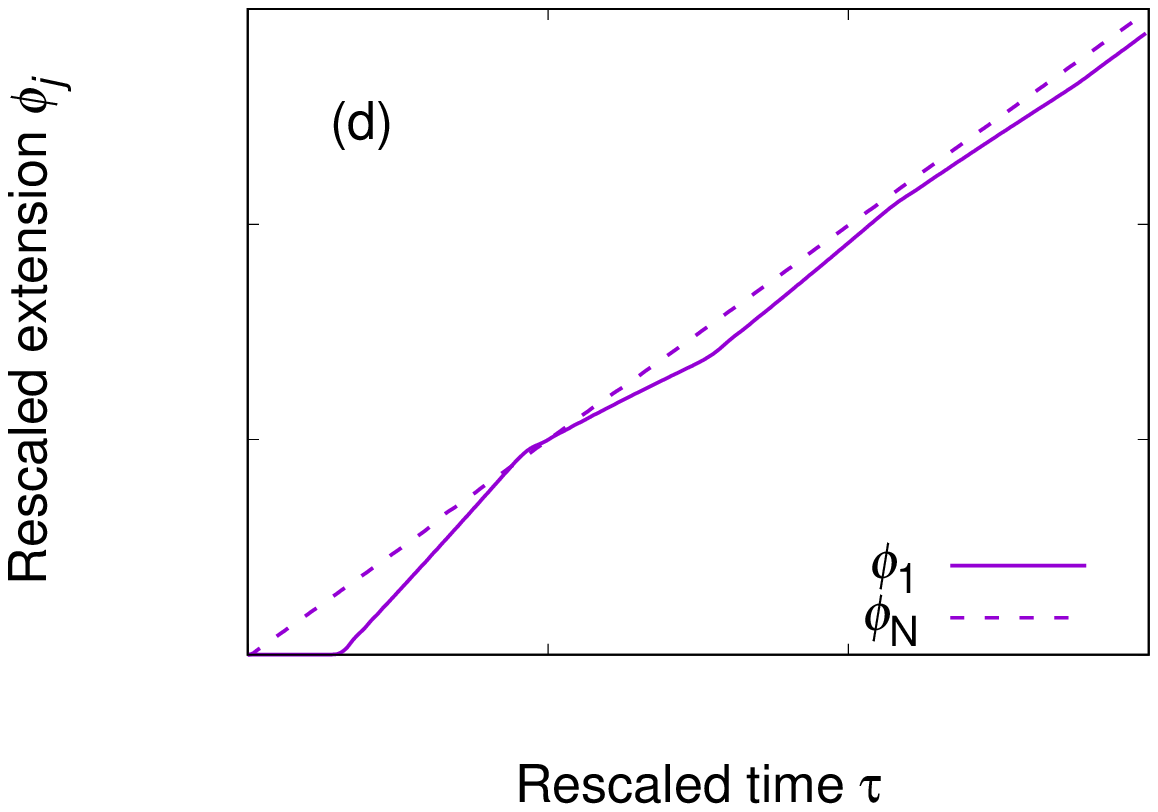}
\caption{(a) Comparison between numerical results (points) and
the approximate solutions in Eqs.~\eqref{eq:extension}--\eqref{eq:phi1}
(lines),
when $\gamma=0$. The data points are obtained from the RK4 method with $N=30$
and $\beta = 1$.
Note that the extension of an intermediate spring such as $\phi_{15}$ lies
between $\phi_1$ and $\phi_{30}$ all the time. (b) Rescaled mechanical energy of
the chain [Eq.~\eqref{eq:energy}], numerically obtained with the same set of
parameters. (c) Which part breaks if every spring behaves as in
Eqs.~\eqref{eq:break} and \eqref{eq:rescale2}. Three different values
of the threshold $\zeta_c$ are represented by the horizontal arrows.
For large $\beta$, the corresponding $\zeta_c$ is small,
so it is the $N$th spring that extends to the threshold at the smallest
$\tau$ (bottom arrow). For small $\beta$, the breaking point can be located
closer to the wall
because $\zeta_N$ is still below the threshold while others exceed it
(middle arrow). However, the $N$th spring can break with even smaller
$\beta$ (topmost arrow), which has been called ``anomalous'' breaking.
(d) Effects of friction when $\gamma=10^{-2}$ in Eq.~\eqref{eq:dimless}. The
other parameters are the same as in panel (a). Because of friction, the triangle
wave of $\phi_1$ decays as time goes by, whereas $\phi_N$ is hardly affected.}
\label{fig:phi}
\end{figure}

If $\gamma=0$ and $f(\tau) = \beta \tau$ with a
constant slope $\beta$, Eq.~\eqref{eq:conv} yields
\begin{equation}
\psi(\xi,\tau) = \beta \tau \xi -\frac{16 \beta N^2}{\pi^3} \sum_{n=0}^\infty
\frac{(-1)^n}{(2n+1)^3}
\sin \left[ \left(n + \frac{1}{2} \right) \frac{\pi \xi}{N} \right]
\sin \left[ \left(n + \frac{1}{2} \right) \frac{\pi \tau}{N} \right].
\label{eq:example}
\end{equation}
A direct way to obtain Eq.~\eqref{eq:example} is to note that $\psi_j = \beta
\tau j$ forms a particular solution for Eq.~\eqref{eq:dimless} when
$\gamma = 0$. For general $\gamma>0$, however, one should employ the method of
Green's functions. The displacement field is obtained by
differentiating Eq.~\eqref{eq:example} with respect to $\xi$
\begin{equation}
\phi(\xi,\tau) = \beta \tau
-\frac{8 \beta N}{\pi^2} \sum_{n=0}^\infty \frac{(-1)^n}{(2n+1)^2}
\sin \left[ \left(n + \frac{1}{2} \right) \frac{\pi \tau}{N} \right]
\cos \left[ \left(n + \frac{1}{2} \right) \frac{\pi \xi}{N} \right].
\label{eq:extension}
\end{equation}
At $\xi=N$, it reduces to
\begin{equation}
\phi_N (\tau) \approx \phi(\xi=N, \tau) = \beta \tau,
\label{eq:phiN}
\end{equation}
which means that the rightmost spring extends linearly in time.
At the other end of the chain, i.e., at $\xi=0$,
the Fourier series on the right-hand side (RHS) of Eq.~\eqref{eq:extension}
describes a triangle wave of period $4N$ and amplitude $\beta N$, which
behaves as $-\beta \tau$ between $\tau=0$ and $N$.
Combining this result with the first term on the RHS of
Eq.~\eqref{eq:extension},
one can see that the extension of the leftmost spring fastened to the
wall is described by the following piecewise linear function:
\begin{equation}
\phi_1(\tau) \approx \left\{
\begin{array}{cl}
0 & \text{~~~if~~} 0 \le \tau < N\\
2 \beta \left[ \tau - (2\nu+1) N \right] & \text{~~~if~~}
(4\nu+1)N \le \tau < (4\nu+3)N\\
4 \beta (\nu+1) N & \text{~~~if~~} (4\nu+3)N \le \tau < (4\nu+5)N,
\end{array}
\right.
\label{eq:phi1}
\end{equation}
where $\nu = 0,1,2,\ldots$. It is plausible that the spring remains
at rest when $0 \le \tau < N$ because it takes time for the external
perturbation to be transferred through $N$ intermediate springs. However,
the subsequent motion is not so self-evident: The spring
suddenly begins to expand twice as fast as the rightmost one
until the expansion stops abruptly at $\tau=2N$, and
this pattern continues periodically.
Application of Eq.~\eqref{eq:sum} actually shows that \emph{every} spring has
such discontinuity in the time derivative of $\phi_j$ except for $j=N$:
From the shape of $\partial G/\partial \xi$ in
Fig.~\ref{fig:psiu}(b), it is easily seen that
\begin{equation}
\phi_j(\tau) \approx \left\{
\begin{array}{cl}
0 & \text{~~~if~~} 0 \le \tau < N-j\\
\beta(\tau+j-N) & \text{~~~if~~} (4\nu+1) N-j \le \tau < (4\nu+1) N+j\\
2\beta[\tau -(2\nu+1) N] & \text{~~~if~~} (4\nu+1) N+j \le \tau < (4\nu+3) N-j\\
\beta(\tau-j+N) & \text{~~~if~~} (4\nu+3) N-j \le \tau < (4\nu+3) N+j\\
4 \beta (\nu+1) N & \text{~~~if~~} (4\nu+3) N+j \le \tau < (4\nu+5) N-j,
\end{array}
\right.
\label{eq:phij}
\end{equation}
with $\nu = 0,1,2,\ldots$ [see, e.g., $\phi_{N/2}$ in Fig.~\ref{fig:phi}(a)].
Note that Eq.~\eqref{eq:phij} is directly proportional to $\beta$. It is because
Eq.~\eqref{eq:dimless} is linear and thus invariant under rescaling every
$\psi_j$ and $\beta$ by a common factor $\lambda > 0$
\begin{equation}
\lambda \frac{d^2}{d\tau^2} \psi_j = \lambda (\psi_{j-1} - 2\psi_{j} +
\psi_{j+1}) -2\gamma \lambda \frac{d}{d\tau} \psi_j
+ \lambda \beta \tau \delta_{j,N}.
\label{eq:rescale}
\end{equation}
In words, the sole effect of choosing a different value for $\beta$ is to
change the overall length scale. No matter how slowly the end of the
harmonic chain is pulled, the periodic discontinuity will not disappear.
Note also that Eqs.~\eqref{eq:phiN} and \eqref{eq:phi1} provide envelopes
for every $\phi_j$ in between [Eq.~\eqref{eq:phij}],
although the lines can sometimes coincide.
As demonstrated in Fig.~\ref{fig:phi}(a), the analytic predictions
of Eq.~\eqref{eq:phij} are well substantiated by direct numerical integration of
Eq.~\eqref{eq:dimless}. One may also check the mechanical energy per particle
\begin{equation}
\varepsilon = \frac{1}{N} \sum_{j=1}^N \frac{1}{2} \left( \frac{1}{N}
\frac{d\psi_j}{d\tau} \right)^2 + \frac{1}{N} \sum_{j=1}^N \frac{1}{2} \left(
\frac{\psi_j - \psi_{j-1}}{N} \right)^2,
\label{eq:energy}
\end{equation}
where the first and second terms represent the kinetic and potential parts,
respectively [Fig.~\ref{fig:phi}(b)]. The factor of $1/N$
inside each pair of parentheses is due to the fact that $\psi_j \sim O(N)$ [see
Fig.~\ref{fig:phi}(a)]. The kinetic part of Eq.~\eqref{eq:energy} turns out to
be a periodic function of $\tau$ with a period of $4N$. The potential energy, on
the other hand, keeps increasing as a quadratic function of $\tau$.

Recall the assumption that every spring obeys Hooke's law up to some threshold
$\phi_c >0$, above which the spring
breaks.\cite{heald1996string,caplan2004ye,shima2014analytic} The restoring
force of the $j$th spring is thus written as
\begin{equation}
f^{\text{res}}_j = \left\{ \begin{array}{cl}
-K \phi_j & \text{if~~} |\phi_j| < \phi_c\\
0 & \text{otherwise (i.e., broken)}.
\end{array} \right.
\label{eq:break}
\end{equation}
Then, the above calculation implies that
the magnitude of $\beta$ is an important factor to determine which spring
breaks. It is related to the fact that a different value of $\beta$ just
rescales every $\phi_j$ with exactly the same factor.
If $\lambda = \beta^{-1}$ in Eq.~\eqref{eq:rescale} and
$\zeta_j \equiv \beta^{-1} \phi_j$, it is a harmonic chain defined by
\begin{equation}
\frac{d^2}{d\tau^2} \zeta_j = (\zeta_{j-1} - 2\zeta_{j} +
\zeta_{j+1}) -2\gamma \frac{d}{d\tau} \zeta_j + \tau \delta_{j,N},
\label{eq:rescale2}
\end{equation}
in which the threshold of a spring becomes $\zeta_c \equiv \phi_c/\beta$.
Large jerkiness therefore maps to a low threshold in this derived system.
As illustrated in Fig.~\ref{fig:phi}(c), the $N$th spring will break when
$\beta$ is large because it is the earliest one that extends to $\zeta_c$.
Conversely, small $\beta$ can break a spring close to the wall. Precisely
speaking, one can only specify the range of springs to break in the latter case.
According to Eq.~\eqref{eq:phij}, all the springs between $\xi=0$ and $\xi^\ast
(\le N)$ are the most extended ones in this chain when $\tau = 3N-\xi^\ast$ (mod
$4N$). In theory, therefore, it is possible to break every spring all at once by
choosing a suitable value of $\beta$ so that every $\phi_j$ reaches the
threshold $\phi_c$ at the same time, which may happen at $\tau = 2N$.
In practice, however, this would mean that the breaking point becomes
very sensitive to experimental noise and mechanical defects.

Another point of Fig.~\ref{fig:phi}(c) is that
one can break the $N$th spring by pulling the end even \emph{more} slowly,
which proves the existence of ``anomalous''
breaking in this system.\cite{heald1996string,caplan2004ye,shima2014analytic}
If this anomaly is hardly observed, the reason could be that
friction is not negligible in any experimental situation.\cite{heald1996string}
If $\gamma$ is positive yet so small that
only the second summation contributes in Eq.~\eqref{eq:green},
the triangle wave in $\phi_1$ will gradually decay as indicated
by $e^{-\gamma \tau}$ in the summand.
On the other hand, it is reasonable to guess that $\phi_N$ will not
experience any notable change, considering that $\partial G / \partial \xi$
containing the friction term identically vanishes at $\xi = N$ due to the
boundary condition.
Consequently, $\phi_1$ is expected to lie below $\phi_N$ in the long run
[Fig.~\ref{fig:phi}(d)]. It implies that the anomaly can indeed be diminished by
friction, but the price is that it also becomes hard to locate the breaking
point close to the wall. For sufficiently large $\gamma$, the one that
breaks will always be the $N$th spring where the force is acting.

\section{Discussion and Conclusion}
\label{sec:conclusion}

To summarize, I have investigated the dynamics of a harmonic chain which is
anchored to a wall at one end and subject to external force at the other end.
By using the method of Green's functions, one can calculate the response of the
system to a general time-varying force.
A simple expression is obtained when the system becomes a continuous
medium composed of a large number of beads and springs [Eq.~\eqref{eq:sum}].
With the simple failure behavior assumed in Eq.~\eqref{eq:break},
anomalous breaking\cite{heald1996string,caplan2004ye,shima2014analytic}
is still a theoretical possibility in this many-body system when
driven by a ramp force $F \propto t$. A nontrivial difference
from the common pull-or-jerk experiment is that
every spring except the last one
exhibits distinct stop-and-go behavior in its extension $\phi_j$
[compare Eq.~\eqref{eq:heald} and Eq.~\eqref{eq:phij}].
It implies that it roughly takes $\Delta t \sim
O(\omega_0^{-1})$ for the external perturbation to travel across a spring, and
this is a fast process compared to system-wide dynamics when $N$ is large.
When one talks about the pull-or-jerk experiment in the context of Newton's law
of inertia, the precise meaning is that $\Delta t \propto m^{1/2}$.
If time is not enough to send an amount of energy across the chain,
therefore, it will be the rightmost spring that breaks. In other words, there
are two competing time scales: One is the intrinsic time scale of the chain,
and the other is that of the driving force. The point is that the experiment
should be understood in terms of these time scales of forced oscillation, in
addition to the law of inertia.

From a technical point of view, the chain is described by a set of coupled,
linear, ordinary differential equations. Although it looks much more difficult
than the one-body counterpart as in Fig.~\ref{fig:ball}, the problem can readily
be handled by standard techniques such as separation of variables and
Green's functions.\cite{boas2006mathematical}
It is instructive to check the validity of the analytic
solution by performing numerical simulations, e.g., with the RK4 method
as we have done throughout this work.
Figure~\ref{fig:phi}(a) has already shown consistency between the analytic
and numerical approaches, but the agreement is actually striking in every
detail, as demonstrated in Fig.~\ref{fig:compare}.
For reference, a sample Python code is provided in the
Appendix.\cite{newman2013computational}
Readers are also encouraged to extend the model to inhomogeneous or anharmonic
cases to incorporate more realistic aspects of the chain
dynamics.\cite{paturej2011polymer}

\begin{acknowledgments}
S.K.B. gratefully acknowledges discussions with Julian Lee, Hang-Hyun Jo, and
Hiroyuki Shima. This work was supported by a research grant of Pukyong National
University (2016).
\end{acknowledgments}


\appendix
\section{Sample code}

Here, I present a Python code to simulate the dynamics of $N=30$ beads with the
RK4 method.\cite{newman2013computational} Note that it includes $y_0 = 0$
explicitly because the index of an array begins from zero by default. The
array \verb#r# contains both the displacement and velocity of every bead in such
a way that its elements \verb#r[2*j]# and \verb#r[2*j+1]# correspond to $y_j$
and $\frac{d}{dt} y_j$, respectively.

\lstset{language=Python}
\begin{lstlisting}
from __future__ import print_function,division # for Python 2
from numpy import empty,array,zeros

def drive(t):
    return beta*t

def f(r,t):
    rdot = empty(2*N1, float)
    for i in range(0, 2*N1, 2):  # time derivative of displacement
        rdot[i] = r[i+1]
    for i in range(1, 2*N1, 2):  # time derivative of velocity
        if i==1:
            accel = 0.
        elif i>1 and i<2*N1-1:
            accel = r[i-3] - 2*r[i-1] + r[i+1] - gamma*r[i]
        elif i==2*N1-1:
            accel = r[i-3] - r[i-1] - gamma*r[i] + drive(t)
        rdot[i] = accel
    return rdot

start, end = 0., 150.            # time domain
max_step = 15000                 # number of time steps
h = (end - start) / max_step     # time increment
N1 = 31    # number of beads including the zeroth one (j=0)
beta = 1.  # increasing rate of the driving force
gamma = 0. # friction coefficient
r = zeros(2*N1, float)
for step in range(max_step):
    t =  h*step
    k1 = h*f(r,t)
    k2 = h*f(r+0.5*k1,t+0.5*h)
    k3 = h*f(r+0.5*k2,t+0.5*h)
    k4 = h*f(r+0.5*k3,t+h)
    r += (k1+2*k2+2*k3+k4)/6
\end{lstlisting}

\newpage

\begin{figure}
\includegraphics[width=0.45\textwidth]{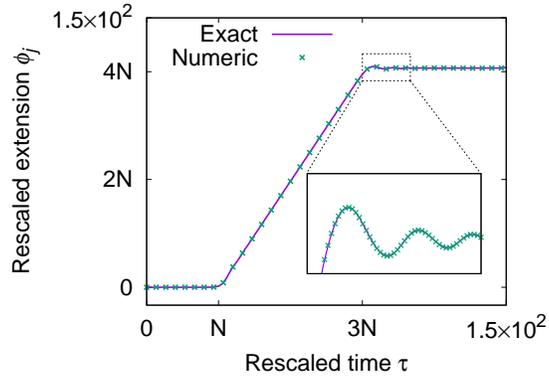}
\caption{Comparison between the analytic solution [Eq.~\eqref{eq:exactconv}] and
the numerical result obtained by the RK4 method when $N=30$, $\gamma=0$, and
$f(\tau) = \tau$. Inset: Zoomed-in view of the small oscillatory part. The
numerical data points are exactly on top of the analytic solution.}
\label{fig:compare}
\end{figure}
\end{document}